\documentclass[12pt]{article}
\usepackage{amssymb}
\usepackage{amsmath}
\usepackage{epsfig}
\usepackage{float}
\newcommand{\be}{\begin{equation}}
\newcommand{\ee}{\end{equation}}
\newcommand{\bea}{\begin{eqnarray}}
\newcommand{\eea}{\end{eqnarray}}
\begin{document}

\begin{center}
{\bf MAJORANA NEUTRINO MIXING }

\end{center}

\begin{center}
S. M. Bilenky
\end{center}

\begin{center}
{\em  Joint Institute
for Nuclear Research, Dubna, R-141980, Russia, and\\
SISSA,via Beirut 2-4, I-34014 Trieste, Italy.}
\end{center}

\abstract{The most natural see-saw explanation of the smallness of the neutrino masses is 
based on the assumption that total lepton number is violated at a large scale and 
neutrinos with definite masses are Majorana particles.
In this review we consider in details  difference
between Dirac and Majorana neutrino mixing and 
possibilities of  revealing Majorana nature of neutrinos with definite masses.}

\section{Introduction}
Phenomenon of neutrino oscillations, discovered in the Super Kamiokande \cite{SK},
SNO \cite{SNO}, KamLAND  \cite{Kamland}, K2K   \cite{K2K} and other neutrino experiments
\cite{Cl,Gallex,Sage,SKsol}  is one of
the most important
signature of a new, beyond the Standard Model physics.

Investigation of neutrino oscillations is based on the following
assumptions:
\begin{enumerate}
\item Neutrino interaction is given by the standard CC and NC
Lagrangians
\be
\mathcal{L}_{I}^{\mathrm{CC}} = - \frac{g}{2\sqrt{2}} \,
j^{\mathrm{CC}}_{\alpha} \, W^{\alpha} + \mathrm{h.c.};~~
\mathcal{L}_{I}^{\mathrm{NC}} = - \frac{g}{2\cos\theta_{W}} \,
j^{\mathrm{NC}}_{\alpha} \, Z^{\alpha}\label{1} 
\ee
Here $g$ is the $SU(2)$ gauge coupling, $\theta_{W}$ is the weak
angle and
\be 
j^{\mathrm{CC}}_{\alpha} =2 \sum_{l=e,\mu,\tau} \bar \nu_{lL}
\gamma_{\alpha}l_{L};~~ j ^{\mathrm{NC}}_{\alpha}
=\sum_{l=e,\mu,\tau} \bar \nu_{lL}\gamma_{\alpha}\nu_{lL}\label{2}
\ee
are the leptonic charged current and the neutrino neutral current.

\item In the total Lagrangian a neutrino mass term, source of neutrino
masses and  mixing, enters.

\end{enumerate}

For neutrinos, particles with equal to zero electric charges, 
two types of  mass terms are possible (see reviews \cite{BilPet,BilGG}):

I. Dirac mass term
\be 
\mathcal{L}^{\mathrm{D}}=- \bar \nu_{R}\,M^{\mathrm{D}} \nu_{L}
+\mathrm{h.c.} \label{3} \ee
Here $M^{\mathrm{D}} $ is a non diagonal complex matrix and

\bea \nu_{L}=\left(
\begin{array}{c}
\nu_{e L}\\
\nu_{\mu L}\\
\nu_{\tau L}\\
\end{array}
\right);~~~ \nu_{R}=\left(
\begin{array}{c}
\nu_{e R}\\
\nu_{\mu R}\\
\nu_{\tau R}\\
\end{array}
\right). \label{4}\eea
After the diagonalization of the matrix $M^{\mathrm{D}}$
 the mass term (\ref{3}) takes the standard form
\be 
\mathcal{L}^{\mathrm{D}}=- \sum^{3}_{i=1}m_{i}\,\bar
\nu_{i}\nu_{i}.\label{5} \ee
and flavor field
$\nu_{lL}(x)$ is given by 
\be
\nu_{l L}(x) = \sum^{3}_{i=1} U_{{l}i} ~ \nu_{iL}(x). \label{6}
\ee
Here $\nu_{i}(x)$ is the field of neutrino with mass $m_{i}$ and $U$ is an unitary matrix.
Thus, in general case of non diagonal matrix $M^{\mathrm{D}}$ flavor fields
$\nu_{lL}(x)$  in charged  and neutral  currents (\ref{2}) are "mixed fields".

In the case of the Dirac mass term  the total Lagrangian is invariant
under global gauge transformations
\be \nu_{lL}(x) \to e^{i\,\alpha}\,\nu_{lL}(x);~~ \nu_{lR}(x)\to
e^{i\,\alpha}\,\nu_{lR}(x);~~ l(x)\to e^{i\,\alpha}\,l(x),\label{7}\ee
where $\alpha$ is an arbitrary constant phase. This  invariance
means that the total lepton number $ L= L_{e}+L_{\mu}+L_{\tau}$ is
conserved and $\nu_{i}(x)$ are four-component fields of neutrinos
($L=1$) and antineutrinos ($L=-1$).

II.  Majorana mass term
\be \mathcal{L}^{\mathrm{M}}=-\frac{1}{2}\,
\overline{(\nu'_{L})^{c}}\,M^{\mathrm{M}} \nu'_{L} +\mathrm{h.c.} 
\label{8} \ee
Here
\be
(\nu'_{L})^{c}= C \bar\nu'^{T}_{L}\,~~\overline{(\nu'_{L})^{c}}= -\nu'^{T}_{L}\, C^{-1}\label{9}
\ee
($C$ is the matrix of the charge
conjugation), 
$M^{\mathrm{M}}$ is a non diagonal, complex matrix and
\bea 
\nu'_{l}=
\left(
\begin{array}{c}
\nu_{eL}\\
\nu_{\mu L}\\
\nu_{\tau L}\\
\vdots\\
\end{array}
\right).\label{10}
\eea
In the column $\nu'_{L}$ in addition to three
flavor fields $\nu_{lL}$ ($l=e,\mu,\tau$) could also be other fields. These fields 
do not enter into the Lagrangian   (\ref{1})  and corresponding particles have no standard electroweak interaction. Such additional fields 
are called sterile.

In the case of the Majorana mass term the total lepton number is
violated and neutrinos with definite masses $\nu_{i}$ are Majorana
particles. We will consider this case in some details.

The Fermi-Dirac statistics of neutrino fields requires that 
\be
(M^{\mathrm{M}})^{T}= M^{\mathrm{M}}. 
\label{11}
\ee
Symmetrical matrix can be diagonalized with the help of an unitary
matrix:
\be 
M^{\mathrm{M}} = (U^{+})^{T}\,m\, U^{+}, 
\label{12} \ee
 where
$U^{\dag}\,U=1$ and $m_{ik}=m_{i}\,\delta_{ik};~~m_{i}>0$.

From (\ref{9}) and (\ref{12}) we find
\be \mathcal{L}^{\mathrm{M}}=-\frac{1}{2}\,\bar\nu^{\rm{M}}m
\nu^{\rm{M}}= -\frac{1}{2}\,\sum_{i}m_{i}\,\bar
\nu_{i}\nu_{i}.\label{13} \ee
Here
\bea \nu^{\rm{M}}= U^{\dagger}\,\nu'_{L}+ (U^{\dagger}\,\nu'_{L})^{c}=
\left(
\begin{array}{c}
\nu_{1}\\
\nu_{2}\\
\nu_{3}\\
\vdots\\
\end{array}
\right)\label{14}\eea
The field $\nu_{i}(x)$ is the field of neutrino with mass $m_{i}$.
From (\ref{14}) it follows that fields $\nu_{i}(x)$ satisfy {\em
Majorana condition}
\be 
\nu^{c}_{i}(x)=\nu_{i}(x),\label{15}\ee
where
$\nu_{i}^{c}(x)=C\,\bar\nu^{T}_{i}(x)$.

From this condition it follows that
\be\nu_{i}(x)=\int N_{p}\,
(e^{-ipx}\,u^{r}(p)\,a^{i}_{r}(p)+e^{ipx}\,v^{r}(p)\,a^{i \dagger}_{r}(p))\,d^{3}p
\label{16}\ee
Here $N_{p}=\frac{1}{(2\pi)^{3/2}}\frac{1}{\sqrt{2p^{0}}}$ and
$a^{i}_{r}(p)$ ( $a^{i \dagger}_{r}(p)$) is the   operator of absorption
(creation) of a Majorana neutrino with momentum $p$, helicity $r$ and mass $m_{i}$.
Thus, Majorana  neutrinos and antineutrinos are {\em identical }.

From (\ref{14}) we find that  flavor fields $\nu_{lL}(x)$ are mixed fields
\be\nu_{l L}(x) = \sum_{i} U_{{l}i} ~ \nu_{iL}(x);~~l=e,\mu,\tau,
\label{17} \ee
where $\nu_{i}(x)$ is the field of the Majorana neutrino with mass $m_{i}$.

If $n_{s}$ sterile fields $\nu_{s_{a}L}$ enter
into the mass term, the number of the Majorana fields $\nu_{i}(x)$ is equal to
$(3+n_{s})$, $U$ is $(3+n_{s})\times (3+n_{s})$ unitary matrix and in
addition to (\ref{17}) we have
\be\nu_{s_{a}L}(x) = \sum_{i} U_{s_{a} i} ~ \nu_{iL}(x);~~a=1,....n_{s}. \label{18}
\ee
So called Majorana and Dirac mass term 
\be 
\mathcal{L}^{\mathrm{M+D}}=
-\frac{1}{2}\,
\overline{(\nu_{L})^{c}}\,M^{\mathrm{M}}_{L} \nu_{L}
- \bar \nu_{R}\,M^{\mathrm{D}} \nu_{L}
-\frac{1}{2}\,
\bar{\nu_{R}}\,M^{\mathrm{M}}_{R} (\nu_{R})^{c}
+\mathrm{h.c.} 
\label{19} \ee
is of special interest. In (\ref{19}) $M^{\mathrm{M}}_{L,R}$ are $3\times3$ Majorana 
symmetrical 
complex matrices and 
$M^{\mathrm{D}}$ is a $3\times3$ Dirac complex matrix. The mass term  $\mathcal{L}^{\mathrm{M+D}}$ can be presented in the form
\be 
\mathcal{L}^{\mathrm{M+D}}=-\frac{1}{2}\,
\overline{(\nu'_{L})^{c}}\,M^{\mathrm{M+D}}\, \nu'_{L} +\mathrm{h.c.} ,
\label{20} 
\ee
where
\bea \nu'_{L}=\left(
\begin{array}{c}
\nu_{ L}\\
(\nu_{R})^{c}\\
\end{array}
\right);~~~ M^{\mathrm{M+D}}=\left(
\begin{array}{cc}
M^{\mathrm{M}}_{L} & (M^{\mathrm{D}})^{T}\\
M^{\mathrm{D}} & M^{\mathrm{M}}_{R}\\
\end{array}
\right). \label{21}\eea
After the diagonalization of the mass term $\mathcal{L}^{\mathrm{M+D}}$ we will find
\be
\nu_{l L}(x) = \sum^{6}_{i=1} U_{{l}i} ~ \nu_{iL}(x);~~
(\nu_{lR}(x))^{c} = \sum^{6}_{i=1} U_{\bar l  i} ~ \nu_{iL}(x);~~l=e,\mu,\tau. \label{22}
\ee

Existing neutrino oscillation data, with the exception of the 
LSND  data \cite{LSND}\footnote{
The LSND result is going to be checked by the MiniBooNE experiment at the Fermilab \cite{Miniboone}}, are in a good agreement with the assumption that the number of 
the massive neutrinos is equal to the 
number of the flavor neutrinos (three) and there are no light sterile neutrinos.
The type of neutrino mass term and, correspondingly, the nature of the massive neutrinos $\nu_{i}$ (Majorana or Dirac?) are unknown. These problems will hopefully be solved in  future experiments.

From neutrino oscillation data only neutrino mass-squared differences 
can be obtained. The mass of the lightest 
neutrino at present is unknown. From the data of the tritium $\beta$-decay experiments 
Mainz \cite{Mainz}
and Troitsk \cite{Troitsk} it was found
\footnote{ Small neutrino mass-squared differences can not be resolved in tritium experiments. 
Thus, $m_{\beta}\simeq m_{0}$, where $ m_{0}$ is common neutrino mass.}
\be
 m_{\beta}\leq 2.3~\rm{eV}.
\label{23}
\ee
From cosmological data for the sum of neutrino masses the upper bounds in the range
\be
\sum_{i} m_{i}\leq (0.4-1.7)~\rm{eV}.
\label{24}
\ee
were inferred (see \cite{Teg2}).

These data together with neutrino oscillation data, which we will discuss in the next section,
 tell us that neutrino masses are different from zero and much smaller then masses of quarks and leptons. For example, for the particles of the third 
generation we have: 
$$m_{t}\simeq 174.3 \,~\rm{GeV}, m_{b}\simeq 4.6 \,\rm{ GeV}, m_{\tau}\simeq 1.78 \,\rm{ GeV}, 
m_{3}\lesssim 2\cdot  10^{-9} \,\rm{ GeV}$$
 Thus, within one generation masses of quarks and lepton differ by
one-two orders of magnitude. Neutrino mass is about 9-11 orders of magnitude smaller than masses of quarks and lepton.

The most plausible mechanism of the generation of neutrino masses which are much smaller than the masses of 
quarks and leptons
is  the see-saw mechanism \cite{see-saw}. 
In order to explain an idea of this mechanism, 
let us consider the case of one generation (say. first). Assume 
that the Dirac mass term 
\be 
\mathcal{L}^{\mathrm{D}}=-m_{D}\,\bar \nu_{eR}\,\nu_{eL}
+\mathrm{h.c.} \label{25} 
\ee
 is generated  by the standard Higgs mechanism via spontaneous 
symmetry breaking. 
Masses of particles, generated by this mechanism, are proportional to the constant 
$v=(\sqrt{2}\,G_{F})^{-\frac{1}{2}}\simeq 250 $ GeV, which characterizes the scale of the electroweak symmetry breaking.
It is natural to expect 
 that $m^{\mathrm{D}}$ is of the same order of magnitude as masses of $u$, $d$ quarks and electron. Experimental data tell us, however, that neutrino mass is much smaller than masses of 
quarks and leptons.
Thus, an additional mechanism which leads to suppression of the neutrino mass is needed. 

Let us assume that 
in addition to (\ref{25}) a lepton number violating 
right-handed Majorana mass term
\be 
\mathcal{L_{R}}^{\mathrm{M}}=-\frac{1}{2} \,M_{R} 
\overline{(\nu_{eR})}\,(\nu_{eR})^{c} +\mathrm{h.c.} 
\label{26} 
\ee
is generated by some mechanism.\footnote{Let us stress 
that such mass term is allowed only for neutrinos.  
Because of the conservation of the electric charge, Majorana mass terms for quarks and leptons are not allowed.}  
The total mass term is of the Majorana and Dirac type
\be 
\mathcal{L}^{\mathrm{M+D}}=-\frac{1}{2}\,
\overline{(\nu'_{L})^{c}}\,M^{\mathrm{M+D}}\, \nu'_{L} +\mathrm{h.c.} ,
\label{27} 
\ee
where
\bea \nu'_{L}=\left(
\begin{array}{c}
\nu_{e L}\\
(\nu_{e R})^{c}\\
\end{array}
\right);~~~ M^{\mathrm{M+D}}=\left(
\begin{array}{cc}
0& m_{D}\\
m_{D}&M_{R}\\
\end{array}
\right). \label{28}\eea
After the standard diagonalization of the mass term (\ref{27}) we find
\be 
\mathcal{L}^{\mathrm{M+D}}= -\frac{1}{2}\,\sum_{i=1,2}m_{i}\,\bar
\nu_{i}\nu_{i}.\label{29} 
\ee
where $\nu_{1,2}$ 
are Majorana fields. Neutrino masses  $m_{i}$ are given by the relation 
\be
m_{1,2}= \frac{1}{2}\,\left| M_{R} \mp \sqrt{M_{R}^{2}+ 4\,m_{d}^{2}}\right |
\label{30}
\ee
For  $\nu_{eL}$  and 
$(\nu_{eR})^{c}$ we find the mixing relations
\bea
\nu_{eL} &=& -i\,\cos \theta \,
\nu_{1L} +\sin \theta\, \nu_{2L}\nonumber\\ 
(\nu_{eR})^{c} &=& i\,\sin \theta \,\nu_{1L} +\cos \theta\, \nu_{2L}, 
\label{31}
\eea
where 
mixing angle $\theta$ is given by 
\be
\tan 2\,\theta = \frac{2\,m_{D}}{M_{R}}.
\label{32}
\ee
Let us assume now that 
the mass $ M_{R}$ which characterize 
the scale of the violation of the lepton number 
is much larger than $ m_{D}$. 
From (\ref{30}) and  (\ref{32}) we obtain in this case
\be
m_{1}\simeq \frac{m_{D}^{2}}{M_{R}}\ll m_{D};~~m_{2}\simeq M_{R};~~
\theta \simeq  \frac{m_{D}}{M_{R}}\ll 1.
\label{33}
\ee
The field of Majorana neutrino and the field of heavy Majorana particle are given by
\be
\nu_{1}\simeq i\,(\nu_{eL}-(\nu_{eL})^{c});~~\nu_{2}\simeq (\nu_{eR}+(\nu_{eR})^{c}).
\label{34}
\ee   
In the case of three generation the standard see-saw symmetrical mass matrix has the form
\bea 
M^{\mathrm{M+D}}=\left(
\begin{array}{cc}
0&  (M^{\mathrm{D}})^{T}\\
M^{\mathrm{D}}&  M_{R}\\
\end{array}
\right), \label{35}\eea
where $(M_{R})^{T}= M_{R}$ and $M^{\mathrm{D}}\ll M_{R}$. By analogy with the case of one generation 
we will choose the unitary matrix $U$ in Eq. (\ref{12}) in the form
\bea 
U=\left(
\begin{array}{cc}
1 & -(M^{\mathrm{D}})^{\dagger}\,(M^{-1}_{R})^{\dagger}\\
M^{-1}_{R}\,M^{\mathrm{D}} & 1\\
\end{array}
\right). \label{36}\eea
From (\ref{35}) and (\ref{36}) in linear approximation we have
\bea 
U^{T}\,M^{\mathrm{M+D}}\,U=\left(
\begin{array}{cc}
(M^{\mathrm{D}})^{T}\,M^{-1}_{R}\, M^{\mathrm{D}}& 0\\
0 & M_{R}\\
\end{array}
\right), \label{37}
\eea
The $3\times 3$ matrix
\be
m_{\nu}=(M^{\mathrm{D}})^{T}\,M^{-1}_{R}\, M^{\mathrm{D}}
\label{38}
\ee
is neutrino mass matrix. 

Thus, 
if see-saw mechanism is the mechanism of the generation of neutrino masses,  in this case:
\begin{itemize}
\item
Neutrinos are Majorana particles with masses which are much smaller than masses of leptons and quarks.

\item Heavy Majorana particles, see-saw partners  of light Majorana neutrinos, must exist.
CP-violating decays of these particles in the early Universe is considered as a probable  source of the barion asymmetry of the Universe (see review  \cite{Barasym}).

\end{itemize}

\section{Briefly on the status of neutrino oscillations}

From (\ref{6}) and (\ref{17}) for the state of flavor neutrino
$\nu_{l}$ produced in a CC weak process
together with lepton $l^{+}$ we have (see, for example, \cite{BilPet,BilGG})
\be 
|\nu_{l}\rangle
=\sum_{i}U^{*}_{li}\,|\nu_{i}\rangle,
\label{40}
\ee
where $|\nu_{i}\rangle$ is the state of neutrino with mass $m_{i}$,
momentum $\vec{p}$ and energy $E_{i}\simeq p +
\frac{m^{2}_{i}}{2p}$.

In the case of the three-neutrino mixing for the probability of the
transition $\nu_{l} \to \nu_{l'}$ in vacuum (during
time $t$)  we find the following expression 
\be 
{\mathrm P}(\nu_{l} \to \nu_{l'}) =|\,\sum^{3}_{i=1}U_{l'  i} \,~ e^{-
i\,E_{i}t} \,~U_{l i}^*\, |^2 =|\delta_{l'l}+ \sum_{i= 2,3} U_{l'
i} \,~( e^{- i\,\Delta m^2_{1i} \frac {L}{2E}}-1) \,~U_{l i}^*\,
|^2,\label{41}
\ee
Here $\Delta m^2_{ik}=m^2_{k}-m^2_{i}$ and $ L\simeq t$ ($L$ is the
distance between production and detection points).

In the general case the  probability $ {\mathrm P}(\nu_{l} \to \nu_{l'})$
depends on six parameters: two neutrino mass-squared differences
$\Delta m^2_{12}$ and $\Delta m^2_{23}$, 
mixing angles $\theta_{12}$, $\theta_{23}$, $\theta_{13}$ and 
CP violating phase $\delta$ (last four parameters characterize PMNS \cite{BP,MNS}
mixing matrix $U$).
Existing data, however, are well described by simple two-neutrino expressions for transition probabilities, which can be obtained from (\ref{41}) in the leading approximations (see \cite{BilGG}).
This approximation is based on the smallness of two parameters 
\be 
\frac {\Delta m^2_{12}}{\Delta m^2_{23}}\simeq 3.3\cdot 10^{-2};~~~
\sin^{2}\theta_{13}\leq 5\cdot 10^{-2}. \label{42}
\ee
The  value of the parameter $\frac {\Delta m^2_{12}}{\Delta m^2_{23}}$
was obtained from analysis of all neutrino oscillation data. 
The inequality  in (\ref{42})  was obtained from analysis of the data of the reactor CHOOZ experiment 
\cite{Chooz}, in which no indications in favor of neutrino oscillations were found.

If we neglect in the transition probability  (\ref{41}) small terms proportional to 
$\frac{\Delta m^2_{12}} {\Delta m^2_{23}}$ and $\sin^{2}\theta_{13}$, we will find that in the
region of $L/E$ in which $\Delta m^2_{23}\,L/E\gtrsim 1$
(atmospheric and long baseline  neutrino experiments) dominant
transitions  are  $\nu_{\mu} \to \nu_{\tau}$ 
and $\bar\nu_{\mu} \to\bar\nu_{\tau}$ .
For the
probability of $\nu_{\mu}$ ($\bar\nu_{\mu}$) to survive from (\ref{41}) we obtain 
the standard two-neutrino expression
\be
{\mathrm P}(\nu_\mu \to \nu_\mu) = 
{\mathrm P}(\bar\nu_\mu \to \bar\nu_\mu)=
1 - \frac {1}
{2}\,\sin^{2}2\theta_{23}\, (1-\cos \Delta m_{23}^{2}\, \frac {L}
{2E}).\label{43}
\ee
In the case of solar and KamLAND experiments for which 
small $\Delta m_{12}^{2}$ is relevant
 effect of the ``large''
$\Delta m_{23}^{2}$ is averaged. 
For $\nu_{e}$ ($\bar\nu_{e}$) survival probability in vacuum 
(or in matter) the following general expression can be obtained \cite{Schramm}:
\be
{\mathrm P}(\nu_e \to \nu_e) ={\mathrm P}(\bar\nu_e \to \bar\nu_e)=
\sin^{4} \theta_{13} +(1-\sin^{2} \theta_{13})^{2}\,{\mathrm
P}^{(12)}(\nu_e \to \nu_e),\label{44}
\ee
where  
${\mathrm P}^{(12)}(\nu_e \to \nu_e)$ is the two-neutrino $\nu_e$ ($\bar\nu_e$)  survival
probability in vacuum (or in matter)  which depends on 
$\Delta m_{12}^{2}$ and $\sin^{2}\,\theta_{12}$.
In the leading approximation
the probability of reactor $\bar\nu_{e}$ to survive in
vacuum is given by
\be
{\mathrm P}(\bar \nu_e \to \bar\nu_e)
=1-\frac{1}{2}~\sin^{2}2\,\theta_{12}~ (1 - \cos \Delta m_{12}^{2}
\,\frac {L}{2E}),\label{45} 
\ee
where the second term is the sum of (approximately equal) transition probabilities 
$ {\mathrm P}(\bar \nu_{e} \to \bar \nu_{\mu})$ and $ {\mathrm P}(
\bar \nu_{e} \to \bar \nu_{\tau})$.

The probability of solar $\nu_{e}$ to survive in matter in the
leading approximation is given by two-neutrino expression which
depend on $\Delta m_{12}^{2}$, $ \tan^{2}\theta_{12}$ and electron number 
density $\rho_{e}(x)$.

From analysis of the Super Kamiokande atmospheric data for the
parameters $\Delta m_{23}^{2}$ and $ \sin^{2}2\theta_{23}$ the
following 90 \% CL ranges were obtained \cite{SK}
\be 
1.5\cdot 10^{-3}\leq \Delta m^{2}_{23} \leq 3.4\cdot
10^{-3}\rm{eV}^{2};~~ \sin^{2}2 \theta_{23}> 0.92.\label{46}
\ee
From the global analysis of solar and KamLAND data it was found that \cite{SNO}
\be 
\Delta m^{2}_{12} = 8.0^{+0.6}_{-0.4}~10^{-5}~\rm{eV}^{2};~~~
\tan^{2} \theta_{12}= 0.45^{+0.09}_{-0.07}.\label{47}
\ee

\section{Majorana mixing matrix}
An unitary $n \times n$ matrix $U $ is characterized by
$\frac{n(n-1)}{2}$ angles and $\frac{n(n+1)}{2}$ phases. The matrix
$U$ can be presented in the form
\be 
U= S^{\dagger}(\beta)~U^{0}~S(\alpha), \label{48} 
\ee
where  $S(\alpha)$ and $S(\beta)$ are
diagonal phase matrices:
\be 
S_{l'l}(\beta)=e^{i\beta_{l}}\,\delta_{l'l};~~
S_{ik}(\alpha)=e^{i\alpha_{i}}\,\delta_{ik}.
\label{49} 
\ee
Because common phase is unobservable, one of $ 2 n$ phases $\beta_{l}$ and $\alpha_{i}$  can be put
equal to zero. We will choose $\alpha_{n}=0$. 

Let us consider first the case of the Dirac neutrinos. Phases of Dirac
fields are arbitrary unphysical quantities. Thus, in the case of Dirac
neutrinos phase factors $e^{i\beta_{l}}$ and $e^{i\alpha_{i}}$ can
be absorbed, respectively, by  the fields $l(x)$ and $\nu_{i}(x)$.
Therefore
the Dirac mixing matrix is given by
\be 
U^{D} =U^{0}.\label{50} 
\ee
The Dirac mixing matrix is characterized by 
$\frac{n(n-1)}{2}$
angles and $$\frac{n(n+1)}{2}-(2n-1)=\frac{(n-1)(n-2)}{2}$$ physical
phases. In $n=3$ case the mixing matrix is characterized by
three angles and one phase. In the standard parametrization the
matrix $U^{0}$ has the form
\begin{eqnarray}
U^{0} =   \left (
  \begin{array}{ccc}
    c_{12} c_{13} & s_{12} c_{13} & s_{13} e^{-i \delta} \\
    -s_{12} c_{23} - c_{12} s_{23} s_{13} e^{i \delta} & c_{12} c_{23} - s_{12}
    s_{23} s_{13}e^{i \delta} & s_{23} c_{13} \\
    s_{12} s_{23} - c_{12} c_{23} s_{13}e^{i \delta} & -c_{12} s_{23} - s_{12}
    c_{23} s_{13}e^{i \delta} & c_{23} c_{13}
  \end{array}
  \right ). 
\label{51}
\end{eqnarray}
Here $s_{ij} = \sin\theta_{ij}$, $c_{ij} = \cos\theta_{ij}$. 

In the case of the  Majorana neutrinos  only phase factors
$e^{i\beta_{l}}$ can be included in the Dirac leptonic fields
$l(x)$. Neutrino fields $\nu_{i}(x)$ satisfy Majorana condition
(\ref{15}) which fix
phases of the fields
$\nu_{i}(x)$. Thus, in the Majorana case phases $\alpha_{i}$ are
physical quantities.  Majorana mixing matrix has the form
\be 
U^{M} =U^{0}\,S(\alpha)\label{52} 
\ee
In the general $n\times n$ case it is characterized by $\frac{n(n-1)}{2}$
angles  and $$\frac{n(n+1)}{2} -n = \frac{n(n-1)}{2}$$
physical phases.

Let us notice that 
we can include (n-1) Majorana phases  $\alpha_{i}$ into  Majorana fields and introduce new fields $\nu'_{i}=e^{i\alpha_{i}}\,\nu_{i}  $. In this case for the mixing we have 
\be
\nu_{lL}=\sum_{i} U^{M}_{li}\,\nu_{i}= \sum_{i} U^{0}_{li}\,\nu'_{i}.
\label{53} 
\ee
The Majorana condition for the fields $\nu'_{i}$ takes the form  
\be
\nu'^{c}_{i}(x)=e^{-2i \alpha_{i}}\,\nu'_{i}(x).
\label{54}
\ee
Thus, there are two alternative ways of the inclusion of the Majorana phases into mixing relations:
\begin{itemize}
\item
We can choose Majorana condition in the form (\ref{15}). In this case 
Majorana mixing matrix is characterized by $\frac{n(n-1)}{2}$ phases and has the form 
Eq.  (\ref{52}).
\item
We can choose Majorana condition in the form (\ref{54}). In this case Majorana mixing matrix 
is characterized by $\frac{(n-1)(n-2)}{2}$ phases and 
has the same form as Dirac mixing matrix. Other $n-1$
physical phases enter into Majorana conditions.
\end {itemize}
We will demonstrate later that both ways give the same 
physical results.

Up to now we have considered general neutrino mixing. Let us discuss now 
the case of the CP invariance in the lepton sector. In this case we have
\be
V_{CP}~\mathcal{L}_{I}^{\mathrm{CC}}(x)~V^{-1}_{CP} = \mathcal{L}_{I}^{\mathrm{CC}}(x'),
\label{55}
\ee
where $V_{CP}$ is the operator of the CP conjugation and $x'= (x^{0}, -\vec{x})$.
In the case of the neutrino mixing the CC Lagrangian has the form
\be
\mathcal{L}_{I}^{\mathrm{CC}}
=
- \frac{g}{\sqrt{2}}~ \sum_{i,l} U^{*}_{li}~\bar \nu_{iL} \gamma_{\alpha}l_{L}~
W^{\alpha}
- \frac{g}{\sqrt{2}}~ \sum_{i,l} U_{li}~\bar l_{L}\gamma_{\alpha} \nu_{iL}~
W^{\alpha \dagger}
\label{56}
\ee
We will consider 
first the case of the Dirac neutrino fields. The arbitrary CP phase factors of the Dirac 
fields can be chosen to be equal to one. For neutrino fields  we have
\be
V_{CP}\nu_{iL}(x)V^{-1}_{CP}=  \gamma_{0}C\bar \nu_{iL}^{T}(x');~~ 
V_{CP}\bar \nu_{iL}(x)V^{-1}_{CP}=- \nu_{iL}^{T}(x')C^{-1}\gamma_{0}
\label{57}
\ee
Taking into account that $C\gamma_{\alpha}C^{-1}= - \gamma^{T}_{\alpha}$
and $V_{CP} W^{\alpha}(x)V^{-1}_{CP}= \eta^{\alpha}(W^{\alpha})^{\dagger}(x')$ 
($\eta^{\alpha}=(-1,1,1,1))$
we find
\be
V_{CP}\bar \nu_{iL}(x) \gamma_{\alpha}l_{L}(x)W^{\alpha}(x)V^{-1}_{CP}
=\bar l_{L}(x')\gamma_{\alpha} \nu_{iL}(x')
W^{\alpha \dagger}(x')
\label{58}
\ee
From (\ref{55}), (\ref{56}) and (\ref{57}) we conclude that in the case of the CP invariance in the 
lepton sector the Dirac mixing matrix is real:
\be
(U^{D})^{*}= U^{D}  
\label{59}
\ee
For Majorana field $\nu_{i}(x)$ we have
\be
V_{CP}\nu_{i}(x)V^{-1}_{CP}= \eta^{*}_{i}\, \gamma_{0} \nu_{i}(x'),
\label{60}
\ee
where $\eta_{i}$ is the CP parity of the Majorana neutrino with mass $m_{i}$.
This quantity can take the values $\eta_{i}= \pm i$.\footnote{In fact, 
taking into account Majorana condition (\ref{15}), from (\ref{60}) we find
$V_{CP}\nu_{i}(x)V^{-1}_{CP}= -\eta_{i}\,\gamma_{0} \nu_{i}(x'). $
Thus, $\eta_{i}= - \eta^{*}_{i}$ or $\eta^{2}_{i}=-1$.} 
With the help of Eq. (\ref{60}) in the Majorana case we obtain the following relation
\be
V_{CP}\bar \nu_{iL}(x) \gamma_{\alpha}l_{L}(x)W^{\alpha}(x)V^{-1}_{CP}
=\eta_{i}\,\bar l_{L}(x')\gamma_{\alpha} \nu_{iL}(x')
W^{\alpha \dagger}(x')
\label{61}
\ee
From  (\ref{55}),   (\ref{56}) and  (\ref{61}) we  find that in the case of the CP invariance in 
the lepton sector Majorana mixing matrix satisfies the relation \cite{BilNedPet} 
\be
(U^{M}_{li})^{*}= \eta^{*}_{i}\, U^{M}_{li}
\label{62}
\ee
\section{Nature of neutrinos can not be revealed via investigation of neutrino oscillations}
Neutrino oscillations is an interference phenomenon. Investigation of neutrino oscillations 
allow to measure such small values of the neutrino mass-squared differences which  
presumably are not reachable in other experiments.
However, the 
study of neutrino oscillations in vacuum or in matter does not allow to  reveal the nature of neutrinos with definite masses \cite{BilHPet,Schechter, Langacker}. 

Let us consider neutrino
oscillations in vacuum. The probability of the transition $\nu_{l}
\to \nu_{l'}$ can be written in the form
\be 
{\mathrm P}(\nu_{l} \to \nu_{l'})
=|\,\sum_{i}U_{l' i} \,~ e^{- i \Delta m^2_{1i}~ \frac {L} {2E}}
~U^{*}_{li} |^2, 
\label{63} 
\ee 
The Majorana mixing matrix is connected with the Dirac mixing matrix by the relation
(\ref{52}).
From Eq.(\ref{63}) it is obvious that
additional $n-1$ Majorana phases $\alpha_{i}$ drop out from the expression
for the transition probability. 
Thus, neutrino transition probability has the same form in the case of 
Dirac and Majorana neutrinos \cite{BilHPet, Schechter} 
\be {\mathrm
P^{M}}(\nu_{l} \to \nu_{l'})= {\mathrm P^{D}}(\nu_{l} \to \nu_{l'})
\label{64} 
\ee
The investigation  of neutrino
transitions in matter also does not allow us to reveal the nature of
massive neutrinos \cite{Langacker}.  

In fact, in the flavor representation the standard
effective Hamiltonian of neutrino in matter is given by the expression \cite{MSW}
\be
H^{m}_{\nu_{l'};\nu_{l}}(t)=\langle \nu_{l'}|H_{0}|\nu_{l}\rangle +
\sqrt{2}\,G_{F}\,\rho_{e}(t)\,X_{\nu_{l'};\nu_{l}}.
\label{65}
\ee
Here $H_{0}$ is the free Hamiltonian, $\rho_{e}$ is the electron number density and
$X_{\nu_{l'};\nu_{l}}= \delta_{\nu_{l'};\nu_{e}}~\delta_{\nu_{l};\nu_{e}}$.

The Hamiltonian of the effective interaction of neutrino with matter
(the second term of (\ref{65})) is determined by the amplitude of the elastic $\nu_{e}-e$
 scattering
in the forward direction. This term does not depend on neutrino masses and mixing.

The neutrino masses and mixing enter only into   the free
Hamiltonian. 
The state of the flavor neutrino is given by the relation 
\be 
|\nu_{l}\rangle
=\sum_{i}U^{*}_{li}\,|\nu_{i}\rangle,
\label{66}
\ee
where $|\nu_{i}\rangle$ is the eigenstate of the free Hamiltonian 
\be 
H_{0}\,|\nu_{i}\rangle= E_{i}\,\,|\nu_{i}\rangle;~~ 
E_{i}\simeq p +
\frac{m^{2}_{i}}{2p}.
\label{67}
\ee
From
(\ref{66})  and  (\ref{67}) (up to unessential unit matrix) 
 we have
\be 
\langle \nu_{l'}|H_{0}|\nu_{l}\rangle = \sum_{i}
U_{l'i}~\frac{m^{2}_{i}}{2 E}~U^{*}_{li}=
\left(U~\frac{m^{2}}{2
E}~U^{\dagger}\right)_{l'; l}.
\label{68} \ee 
Taking into account (\ref{52}),  we have
\be
U^{M}~\frac{m^{2}}{2 E}~U^{M}=U^{D}~\frac{m^{2}}{2 E}~U^{D\,
\dagger}. \label{69} 
\ee
 Thus, additional Majorana CP phases
$\alpha_{i}$ do not enter into the Hamiltonian of neutrino in
matter. In other words the Hamiltonian of neutrino in matter has the
same form for Dirac and Majorana neutrinos and the 
nature of neutrinos can not be revealed through the investigation of
transitions of neutrinos in matter.

\section{On the equivalence of theories with massless
Dirac and Majorana neutrinos}

All existing data are in perfect agreement with the assumption that neutrino interaction is given by the standard Lagrangians  (\ref{1})  and  (\ref{2}). 
For such an interaction
theories with massless Dirac and Majorana neutrinos are equivalent \cite{Okubo}. 
This theorem is based on the fact that for massless neutrinos left-handed Dirac and Majorana fields are connected by the unitary transformation 
\be
V\,\nu^{\mathrm{D}}_{lL}\, V^{-1}= \nu^{\mathrm{M}}_{lL}.
\label{70} 
\ee

The equivalence of theories with massless Dirac and Majorana neutrinos can be seen in 
the following way. 
From  Majorana condition (\ref{15}) it follows that for the Majorana field
right-handed and
 left-handed components are connected by the relation
\be 
\nu_{iR}(x)=(\nu_{iL}(x))^{c}
\label{71} 
\ee
 In the Dirac case right-handed and left-handed components are independent.  
If $m_{i}= 0$,
the right-handed fields do not enter into the standard Lagrangian.
Hence, there is no way to distinguish Dirac and Majorana neutrinos in this case.

If neutrino masses are equal to zero 
$\nu_{lL}(x)$ in (\ref{1}) and  (\ref{2})
are quantum fields. In this case invariance of the Lagrangian under the global gauge transformations 
\be
l'(x)=e^{i\Lambda_{l}}\,l(x);~~\nu'_{lL}(x)=e^{i\Lambda_{l}}\,\nu_{lL}(x) ~~(l=e, \mu, \tau)
\label{72} 
\ee
takes place  and  
flavor lepton numbers $L_{e}$, $L_{\mu}$ and $L_{\tau}$
are strictly  conserved. 

Thus, in the framework of the standard weak interaction observable effects of the massive Majorana 
neutrinos are proportional to neutrino mass.
 Because neutrino masses are very small, {\em  effects which
allow to reveal the Majorana nature of neutrinos are strongly suppressed}.

\section{Violation of total lepton number in neutrino processes}
In the case of the neutrino mixing the standard Lagrangian of the CC interaction of lepton-neutrino 
pair with   $W$-boson is given by  (\ref{56}).
Two types of neutrino (and antineutrino) processes can be induced by the interaction  (\ref{56}).
In the processes of the first type 
neutrino-production process is  due to one term of the interaction (\ref{56}) and neutrino-detection 
process is  induced by  {\em another} term of the interaction. For example, the first term of  (\ref{56})
provides production of neutrino in the transition 
\be
W^{+}\to l^{+}+ \nu_{i}
\label{73}
\ee
Due to the second term of the Lagrangian  (\ref{56}) neutrino can be absorbed in the 
transition
\be
 \nu_{i}\to  l'^{-} +   W^{+}.
\label{74}
\ee
In the physical processes, based on the transitions  (\ref{73}) and  (\ref{74}), total lepton number is conserved. Neutrino in transitions (\ref{73}) and  (\ref{74}) can be Dirac or Majorana particles.

In the matrix elements of the processes of 
production and absorption of neutrino with momentum $p$ and mass 
$m_{i}$ enter the spinor $ 
\frac{1-\gamma_{5}}{2}~u^{r}(p)$. Taking into account
linear in $\frac{m_{i}}{2E}$ terms,  we have 
\be
\frac{1-\gamma_{5}}{2}~u^{r}(p)= \frac{1-r}{2}~u^{r}(p)+
r\frac{m_{i}}{2E}\gamma^{0}u^{r}(p).
\label{75}
\ee
In neutrino experiments energies of neutrino are much larger than 
neutrino masses. Thus, in transitions (\ref{73}) and (\ref{74}) predominantly left-handed 
neutrinos are produced and absorbed. 

Due to Heisenberg uncertainty relation production and absorption of different $\nu_{i}$ can not 
be resolved in neutrino experiments. As a result together with $ l^{+}$ flavor neutrino $\nu_{l}$ is produced. Lepton  $l'^{-}$ can be produced in a CC weak process by flavor neutrino  $\nu_{l'}$. The state of flavor neutrino is given by the relation  (\ref{40}). All observed neutrino process are based on 
 (\ref{73})  and (\ref{74}).

{\em If neutrinos $\nu_{i}$ are Majorana particles} the second type of neutrino processes are possible. 
For Majorana neutrinos operator $\nu_{iL}(x)$ is the sum of the operators of absorption and creation of neutrinos.
Thus, neutrinos which are produced in transition  (\ref{73}) due to the first term of the Lagrangian 
 (\ref{56}) can be absorbed in the transition
\be
 \nu_{i}\to  l'^{+} +   W^{-}
\label{76}
\ee
due to  the {\em same} term of the Lagrangian. It is obvious that in the chain of the processes 
induced by  (\ref{73}) and  (\ref{76}) the total lepton number is changed by two. In the matrix element of the 
absorption of neutrino enter  spinor $\frac{1-\gamma_{5}}{2}~v^{r}(p)$ 
($v^{r}(p)=C\,(\bar u^{r}(p))^{T}$). Taking into account linear in $\frac{m_{i}}{2E}$ terms we have
\be
\frac{1-\gamma_{5}}{2}~v^{r}(p)= \frac{1+r}{2}~v^{r}(p)+
r\frac{m_{i}}{2E}\gamma^{0}v^{r}(p),
\label{77}
\ee
The chain of the processes induced by the transitions (\ref{73}) and  (\ref{76})
 (for example, $\pi^{+}\to \mu^{+}+ \nu_{i};~~\nu_{i} +N\to e^{+} +X$). 
is, however,  strongly suppressed. In fact, from (\ref{75}) it follows that 
in the neutrino-production process 
mainly left-handed neutrinos are produced. From  (\ref{77}) we see that 
in the cross section of the absorption of such neutrinos small factors  
$(\frac{m_{i}}{2E})^{2}$ enter. The probability of the production of the right-handed neutrinos, which have ``large'' weak absorption cross section, is suppressed by the factors $(\frac{m_{i}}{2E})^{2}$.
Thus, the chain of the processes, induced by the first term of the Lagrangian 
(\ref{56}), in which  $l^{+}$ and $l'^{+}$
are produced, are suppressed with respect to usual neutrino processes, induced by the first and the second terms of the Lagrangian (\ref{56}), by the helicity suppression 
factor not larger than 
$(\frac{m_{3}}{2E})^{2}$. ($m_{3}$ is the mass of the heaviest neutrino)\footnote{ It is clear that the same arguments are applied 
to the second term of the Lagrangian  (\ref{56}). This term induces
the chain of neutrino processes in which $l^{-}$ and $l'^{-}$ are produced.}
Taking into account that neutrino energies in neutrino processes $\gtrsim$ MeV and 
$m_{3}\lesssim$ 2 eV we conclude  that the suppression factor is extremely small
\be
\frac{m^{2}_{3}}{4E^{2}}\leq 10^{-12}. \label{78} 
\ee 
Thus,
it is not possible in foreseeable future to reveal
the neutrino nature in neutrino experiments via the observation of the violation 
of the total lepton number.\footnote{ In 1957 R.Davis \cite{Davis}
made an experiment in which he searched for $^{37}\rm{Ar}$ production
in a process of the interaction of antineutrinos from reactor with $^{37}\rm{Cl}$.
He did not find  $^{37}\rm{Ar}$ in the detector and obtain upper bound on the corresponding 
cross section.
As we discussed before in the case of massive Majorana neutrinos  such process in principle is allowed.
It is suppressed, however, by the extremely small factor (\ref{78}).}

\section{$|\Delta L|=2$ processes with virtual Majorana neutrinos}

We will consider  now
$|\Delta L| =2$ processes with virtual Majorana neutrinos which are induced by the first (or the 
second) term of the Lagrangian (\ref{56}).
Examples of such processes are neutrinoless double $\beta$-decay
($0\nu\beta\beta$-decay) of even-even nuclei
 \be (A,Z) \to (A,Z+2)+e^{-} +e^{-},
\label{79} 
\ee 
the decays 
\be
 K^{+}\to \pi^{-}+\mu^{+} +\mu ^{+};~~~K^{-}\to \pi^{+}+\mu^{-} +e ^{-},
\label{80} \ee 
the process
\be 
\mu{-}+(A,Z) \to (A,Z-2)+e^{+} ,
\label{81} \ee 
etc.
The leptonic part of the operator which give contribution to matrix
elements of the processes of the type (\ref{79}) and (\ref{80}) 
is given by (see \cite{BilPet}) 
\be 
\sum_{i,k}T(\bar
l_{L}(x_{1})\,\gamma_{\alpha}~ 
\langle 0|\,T(\nu_{iL}(x_{1})\,\nu^{T}_{kL}(x_{2}))\, |0 \rangle
\,\gamma^{T}_{\beta} \,\bar l'^{T}_{L}(x_{2}))\,U_{li}\,U_{l'k}
\label{82} 
\ee
Let us consider neutrino propagator. 
From the Majorana condition  (\ref{15}) we have
\be
\nu^{T}_{k}=-
\bar\nu_{k}\,C
\label{83} 
\ee
Taking into account this relation we find
\bea
&\langle 0|\,T(\nu_{iL}(x_{1})\,\nu^{T}_{kL}(x_{2}))\, |0 \rangle=&\nonumber\\
&-\delta_{ik}\,\frac{i}{(2\,\pi)^{4}}\, \frac{1-\gamma_{5}}{2} \int
{e^{-ip(x_{1}-x_{2})}\,\frac{\gamma \,p + m_{i}}{p^{2}-m^{2}_{i}}d^{4}p}\,
\frac{1-\gamma_{5}}{2}\,C&
\label{84} 
\eea
For small neutrino masses  $m^{2}_{i} \ll p^{2}$ and we have 
\bea
&\langle 0|\,T(\nu_{iL}(x_{1})\,\nu^{T}_{kL}(x_{2}))\, |0 \rangle\simeq &\nonumber\\
&-\delta_{ik}\, m_{i}\,\frac{i}{(2\,\pi)^{4}}\, \int
{e^{-ip(x_{1}-x_{2})}\,\frac{1}{p^{2}}d^{4}p}\,
\frac{1-\gamma_{5}}{2}\,C&
\label{85} 
\eea
Thus, the propagator of the left-handed components of the neutrino fields 
is proportional to neutrino mass. In the limit $m_{i} \to 0$ processes of the type 
(\ref{78}), (\ref{79}), (\ref{80}) are forbidden in accordance with  the Dirac-Majorana 
equivalence theorem discussed in the 
section 5.

From (\ref{82}) and (\ref{85}) it follows that the matrix elements of the processes
of the type (\ref{79}) and (\ref{80}) are proportional to\footnote{ It is easy to show that the two ways 
of introduction of  Majorana phases, discussed above, give the same result. In fact, we obtained 
(\ref{86}) assuming that the Majorana condition has the form (\ref{15}). If the Majorana condition has the form (\ref{54}) we have $\nu'^{T}_{k}=-e^{2i\alpha_{i}}\,\bar \nu'_{k}\,C$ and 
$$ m'_{ll'} = \sum_{i}U^{0}_{li}\,U^{0}_{l'i}e^{2i\alpha_{i}}\,m_{i}=
\sum_{i}U_{li}\,U_{l'i}\,m_{i}=m_{ll'}$$} 
\be
 m_{ll'} = \sum_{i}U_{li}\,U_{l'i}\,m_{i}.
 \label{86} 
\ee
Analogously, the matrix elements of the processes of the type (\ref{81}) are proportional to  
$ m^{*}_{ll'}$. Taking into account the unitarity of the mixing matrix, we have
\be
 |m_{ll'}| \leq m_{3},
 \label{87} 
\ee
where $m_{3}$
is  the mass of the heaviest neutrino. If we take into account the data of the Mainz and Troitsk experiments \cite{Mainz,Troitsk} we find  
\be
 |m_{ll'}| \leq 2.3 ~\rm{eV}.
 \label{88} 
\ee
The probabilities of the $|\Delta L|=2$ processes with virtual Majorana neutrinos are 
extremely small.
First, they are the processes of the second
order in the Fermi constant $G_{F}$.
And, second, they are helicity suppressed processes.
In the
probabilities of such processes enter very  small helicity
suppression factor
$\frac{m^{2}_{3}}{<Q^{2}>},$
where $<Q^{2}>$ is an average momentum-transfer squared (typically 
$\gtrsim 10~\rm{MeV}^{2}$).

The sensitivities of the experiments on the search for the processes (\ref{80}),   
(\ref{81}) and other similar processes are much worse than the upper bound (\ref{88}).  
In the latest experiment \cite{Kaulard} on the search for the process
$\mu^{-}\rm{Ti}\to e^{+}\rm{Ca}$ the following upper bound was obtained 
$$\frac{\Gamma(\mu^{-}\rm{Ti}\to e^{+}\rm{Ca})}{\Gamma(\mu^{-}\rm{Ti}\to \rm{all})}
\leq 1.7 \cdot10^{-12}$$. The best upper bound on the probability of the decay
$K^{+}\to \pi^{-}\mu^{+} \mu^{+}$ was reached  in \cite{Appel} :
$$\frac{\Gamma(K^{+}\to \pi^{-}\mu^{+} \mu^{+})}{\Gamma(K^{+}\to \rm{all})}\leq 3\cdot 10^{-9}$$
From these data it was obtained, correspondingly, the following bounds \cite{Vogel} 
$$|m_{\mu e}|\leq 82~ \rm{MeV};~~|m_{\mu \mu}|\leq 4\cdot 10^{4}~ \rm{MeV}$$

The exceptional process, sensitive to the expected Majorana neutrino masses, 
is  neutrinoless double
$\beta$-decay of some even-even nuclei.
Possibilities to use
large targets (in present-day experiments tens of kg, in future
experiments about 1 ton and may be more), to reach small background and high energy resolution
make experiments on the search for this decay an unique source of information about 
the nature of massive neutrinos $\nu_{i}$. In the next section we will consider this process in some details.

\section{Neutrinoless double $\beta$-decay}
\subsection{Probability of the decay. Experimental data}
The standard effective  Hamiltonian of the weak interaction of electron-neutrino pair and 
hadrons is given by
\be 
\mathcal{H}_{I}^{\mathrm{CC}}= \frac{G_{F}}{\sqrt{2}}\,2 \bar
e_{L}\gamma _{\alpha}\nu_{eL}\,~ j^{\alpha} + \mathrm{h.c.}, 
\label{89} 
\ee
where $j^{\alpha}$ is the hadronic charged current.
We will assume the Majorana neutrino mixing
\be 
\nu_{eL} = \sum_{i} U_{ei} \nu_{iL}~, \label{90} 
\ee
where $\nu_{i}$ is the  field of Majorana neutrino with mass $m_{i}$.
The matrix element of $0 \nu \beta\beta$-decay is given by the following expression
\bea
\langle f|~S~|i \rangle=-4\,(\frac{G_{F}}{\sqrt{2}})^{2}\,\frac{i}{(2\,\pi)^{4}}\,N_{p_{1}}\,N_{p_{2}}\,
m_{\beta\beta}\,\bar u(p_{2} )\,\gamma_{\alpha}\,\frac{1-\gamma_{5}}{2}\,\gamma^{T}_{\beta}\,
\bar u^{T}(p_{1} )\cdot \nonumber\\
\int e^{ip_{1} x_{1}+ ip_{2} x_{2}}\,e^{-ip(x_{1}-x_{2})}\frac{1}{p^{2}}\,
\langle \Psi_{f}|T( J^{\alpha}(x_{1}) J^{\beta}(x_{2})|\Psi_{i}\rangle d^{4}x_{1}d^{4}x_{2}d^{4}p
 \label{91}
\eea
Here
\be 
m_{ee} \equiv m_{\beta\beta}= \sum_{i}U^{2}_{ei}\,m_{i}. \label{92} 
\ee
is {\em effective Majorana mass},  $|\Psi_{i}\rangle$ and  $|\Psi_{f}\rangle$
are vectors of the initial and final nuclei, $ N_{p_{i}}=\frac{1}{(2\,\pi)^{3/2}} 
\frac{1}{(\sqrt{2\,p_{i}^{0}}}$ and $J^{\alpha}(x)$ is hadronic CC in Heisenberg representation.
From (\ref{91})  for half-life  of the $0 \nu \beta\beta$-decay 
the following
expression can be obtained (see reviews
\cite{bbreviews,FaesslerSuhonen})
\be
\frac{1}{T^{0\,\nu}_{1/2}(A,Z)}=
|m_{\beta\beta}|^{2}\,|M(A,Z)|^{2}\,G^{0\,\nu}(E_{0},Z). 
\label{93}
 \ee 
Here
$M(A,Z)$ is nuclear matrix element (NME) 
and $G^{0\,\nu}(E_{0},Z)$ is  phase-space factor
($E_{0}$ is the energy release). Let us stress that  
NME is determined only by nuclear properties and strong interaction and does not depend on neutrino
masses and mixing.\footnote{If $\nu_{i}$ are Majorana particles 
the $0\nu\beta\beta$-decay mechanism considered here definitely exists. 
Let us notice that there are different beyond the Standard Model mechanisms of violation of the total lepton number and  $0\nu\beta\beta$-decay : the right-handed currents, 
SUSY with violation of $R-$parity, etc. (see \cite{FaesslerSuhonen})}

There exist at present data of 
many experiments on the search for $0 \nu \beta\beta$-decay (see \cite{bbdata}).  
The most
stringent lower bounds on the half-time of $0\,\nu \beta\,\beta $-
decay were obtained in the recent  CUORICINO \cite{Cuoricino}
and  in the Heidelberg-Moscow
\cite{HM} experiments. \footnote{ An  indication in favor of 
the $0 \nu \beta\beta$-decay of  $^{76} \rm{Ge}$,
found in \cite{Klop}, is going to be checked by the GERDA experiment started at Gran Sasso \cite{Gerda}.}

CUORICINO is a cryogenic experiment on the search for   $0\nu\beta\beta$ decay 
of $^{130} \rm{Te}$. 
An array of 62 $\rm{Te} \rm{O}_{2}$ crystals with total mass 40.7 kg
was placed  in a refrigerator at the temperature $T=8$ mK. Heat capacity is proportional to $T^{3}$ 
and the increase of the temperature at tiny
energy release can be recorded in the experiment. 
$^{130} \rm{Te}$ nuclei has high  natural abundance (33.8\%) and 
relatively large $Q$-value ($Q=2528.8\pm1.3$ keV). The background 
in $0\nu\beta\beta$ region in the CUORICINO experiment was  0.18 counts/keV~kg~year.
No signal in the  region of the $0\nu\beta\beta$ decay of $^{130} \rm{Te}$ 
was found. For the half-life the following lower bound was obtained
in the CUORICINO experiment\cite{Cuoricino}
\be 
T^{0\,\nu}_{1/2}(^{130} \rm{Te})\geq 1.8 \cdot 10^{24}\, \rm{years};~~(90\% CL)
\label{94}
\ee 
Taking into account different calculations of NME  from
(\ref{94}) for the upper bound of the effective Majorana mass the following range
of values  was inferred
\be
|m_{\beta\beta}| \leq
(0.2-1.1)\,~\rm{eV}. ~~~(\rm{CUORICINO}) 
\label{95} 
\ee
In the Heidelberg-Moscow experiment \cite{HM} the $0\nu\beta\beta$ decay of $^{76} \rm{Ge}$ was 
studied. 5 crystals of 86 \% enriched $^{76} \rm{Ge}$ with the total mass 11.5 kg was used. 
In the $0\nu\beta\beta$ range of energies the energy resolution and background were
$4.23 \pm 0.14$ keV and 0.163 counts /kg~keV~year.
For the half-life of $^{76} \rm{Ge}$ the following lower bound was obtained in the experiment:
\be 
T^{0\,\nu}_{1/2}(^{76} \rm{Ge})\geq 1.55 \cdot 10^{25}.
years
\label{96}
\ee
Taking into account uncertainties in NME calculations from this result for the effective 
Majorana mass the following bounds can be obtained 
\be 
|m_{\beta\beta}| \leq
(0.3-1.2)\,~\rm{eV}. ~~~(\rm{Heidelberg-Moscow})
\label{97}
\ee
Several future experiments on the search for  $0\,\nu \beta\,\beta $-decay
are in preparation at present.
Detectors in these experiments will be much larger than in today's experiments (
about 1 ton and even more).
All groups, preparing new experiments,  plan to decrease significantly background and to improve 
energy resolution. 
The aim of the future experiments on the search for  $0\,\nu \beta\,\beta $-decay
is to reach sensitivity 
\be 
|m_{\beta\beta}| \simeq \rm{a~few} ~10^{-2} ~\rm{eV}.
\label{98}
\ee
We will mention several proposals (for more detailed discussion see  \cite{bbfuture}). 

The CUORE experiment \cite{Cuoricino} 
will be based on the same 
technique as the CUORICINO experiment. An array of 988 
of $\rm{Te} \rm{O}_{2}$ crystals with total mass 741 kg will be  used. Significant reduction of the background (up to $10^{-2}- 10^{-3}$ counts/keV~kg ~year) and improvement of the energy resolution ($\simeq$ 5 keV) are planned to be reached.
Expected sensitivity to the effective Majorana mass will be about $(2-3)~10^{-2}$ eV.

Majorana collaboration \cite{Frank} plans to use 500 kg of 86 \% enriched $^{76} \rm{Ge}$. 
It is expected that the granularity of the detector and improved pulse-shape analysis will reduce 
background significantly. The anticipated sensitivity of the Majorana experiment is equal to
\be 
T^{0\,\nu}_{1/2}(^{76} \rm{Ge})\simeq 4 \cdot 10^{27}\,
years
\label{99}
\ee
This corresponds to the following sensitivity to the effective Majorana mass 
\be 
|m_{\beta\beta}| \simeq (2-7) \cdot10^{-2} ~\rm{eV}.
\label{100}
\ee
In the EXO experiment \cite{EXO} $0\nu\beta\beta$-decay of $^{136} \rm{Xe}$ will be searched for. 
In this experiment not only the total energy of two electrons will be measured but also daughter 
nuclei  will be detected.
In the 
$0\nu\beta\beta$-decay of $^{136} \rm{Xe}$ ion 
$^{136} \rm{Ba}^{++}$ is produced. This ion will be neutralized to $^{136} \rm{Ba}^{+}$ and localized.
Then $^{136} \rm{Ba}^{+}$ ion will be optically detected through the irradiation by photons from two 
lasers. It is expected that about $10^{7}$ photons/sec will be emitted by one ion. 
If the method of detection of 
$^{136} \rm{Ba}^{+}$ will be realized the only background in the EXO experiment will come from 
$2\nu\beta\beta$-decay.  Improvement of the energy resolution is crucial for the success of the experiment. With 10 ton enriched $^{136} \rm{Xe}$ TPC detector and $^{136} \rm{Ba}^{+}$  tagging the sensitivity 
\be 
T^{0\,\nu}_{1/2}(^{136} \rm{Xe})\simeq 1 \cdot 10^{28}\,
years
\label{101}
\ee
is expected. This sensitivity corresponds to
\be 
|m_{\beta\beta}| \simeq (1.3-3.7) \cdot10^{-2} ~\rm{eV}.
\label{102}
\ee

\subsection{Effective Majorana mass and neutrino oscillation data}
The effective Majorana mass $m_{\beta\beta}$ is 
determined by the values of neutrino masses $m_{i}$ and elements
$U^{2}_{ei}$. In the standard parametrization we have
\bea
U^{2}_{e1}&=&\cos^{2}  \theta_{13}~\cos^{2} \theta_{12}~e^{2i\alpha_{1}};~     
U^{2}_{e2}=\cos^{2}  \theta_{13}~\sin ^{2}\theta_{12}~e^{2i\alpha_{2}};~\nonumber\\
U^{2}_{e3}&=&\sin^{2} \theta_{13} ~e^{2i\alpha_{3}}.
\label{103} 
\eea
Information about the parameters $\tan^{2} \theta_{12}$ and $\sin^{2} \theta_{13}$ can be 
obtained from the data of solar-KamLAND and CHOOZ experiments (see (\ref{47}) and (\ref{42})).
The phases $\alpha_i$ are unknown.

The values of the neutrino masses are determined by the neutrino mass-squared differences 
$\Delta m^{2}_{23}$ and $\Delta m^{2}_{12}$, which are known from the data of the Super-Kamiokande and solar-KamLAND experiments (see (\ref{46}) and (\ref{47})), the mass of the lightest neutrino and character of the neutrino mass spectrum.

For three neutrino masses two neutrino mass spectra are possible:
\begin{enumerate}
\item Normal spectrum
\be 
m_{1}<  m_{2}    <  m_{3} ;~ \Delta m^{2}_{12}  \ll    \Delta m^{2}_{23}\label{104}
\ee
\newpage
\item Inverted spectrum\footnote{In order to keep for the solar-KamLAND
neutrino mass-squared difference notation $\Delta m^{2}_{12}>0 $, 
neutrino masses are usually labeled differently 
in the cases of normal
and inverted neutrino spectra. In the case of the normal spectrum 
$\Delta m^{2}_{23}>0$ and in the case of  the inverted spectrum 
$\Delta m^{2}_{13}<0$.
Thus, with such notations of the neutrino masses the character of the neutrino mass spectrum is determined by the sign
of atmospheric neutrino mass-squared difference. 
It clear, however, that the sign of of the atmospheric neutrino mass-squared difference has no physical meaning: it is a convention based on the notation 
$\Delta m^{2}_{ik}= m^{2}_{k}- m^{2}_{i}$.
Notice that 
in both cases of neutrino mass spectra for mixing angles the same notations can be used
}
\be 
m_{3}<  m_{1}    <  m_{2} ;~ \Delta m^{2}_{12}  \ll
 | \Delta m^{2}_{13}| \label{105}
\ee
\end{enumerate}
 For neutrino masses in the case of the
normal spectrum we have
\be 
m_{2}=\sqrt{m^{2}_{1}+\Delta m^{2}_{12}},~~
m_{3}=\sqrt{m^{2}_{1}+\Delta m^{2}_{12}+\Delta
m^{2}_{23}}.\label{106}
\ee
In the case of the inverted spectrum neutrino masses are given by
\be 
m_{1}=\sqrt{m^{2}_{3}-\Delta m^{2}_{13}},~~
m_{2}=\sqrt{m^{2}_{3}-\Delta m^{2}_{13}+\Delta
m^{2}_{12}}.\label{107}
\ee
The lightest neutrino mass $m_{1}(m_{3})$ is
at present unknown. Upper bound for the neutrino mass obtained from the data of the tritium
experiments, is given in (\ref{23}).

In the leading approximation 
neutrino transition probabilities 
have the same form for both types of neutrino mass spectra. 
Thus, in order to reveal the type of the neutrino mass spectrum
it is necessary to study small effects beyond the leading approximation.
The size of such effects and possibilities to measure them depend on the value of the parameter
$\sin^{2}\theta_{13}$. We will see that the 
investigation of the neutrinoless double-$\beta$-decay
will allow us to obtain an information about neutrino mass spectrum independently on the value of the small parameter $\sin^{2}\theta_{13}$.

We will consider three standard neutrino mass spectra (see \cite{bbpapers})

I. Hierarchy of neutrino masses 
\be m_{1} \ll m_{2} \ll
m_{3} 
\label{108} 
\ee
In this case neutrino masses  $m_{2,3}$ are determined by neutrino mass-squared differences:
\be
 m_{2}\simeq \sqrt{ \Delta m^{2}_{12}}\simeq 8.9 \cdot 10^{-3}\rm{eV};
m_{3}\simeq  \sqrt{ \Delta m^{2}_{23}}\simeq 4.9 \cdot 10^{-2}\rm{eV}.
\label{109} 
\ee
The lightest neutrino mass satisfies inequality: $m_{1} \ll
\sqrt{\Delta m^{2}_{12}}.$
Neglecting the contribution of $m_{1}$, 
for the effective Majorana mass we obtain the following expression
\be
|m_{\beta\beta}|\simeq\left |\,
 \sin^{2} \theta_{12}\, \sqrt{\Delta m^{2}_{12}} + e^{i\,\alpha_{23}}
 \sin^{2} \theta_{13}\, \sqrt{\Delta m^{2}_{23}}\,\right|,
 \label{110}
\ee
where $\alpha_{23}=\alpha_{3}- \alpha_{2} $.

The first term in Eq.(\ref{110}) is small because of the smallness of
$\sqrt{\Delta m^{2}_{12}}$. Contribution of ``large'' $\sqrt{\Delta
m^{2}_{23}}$ is suppressed by the small factor $\sin^{2} \theta_{13}
$. If we will use the  CHOOZ  bound (\ref{42}) the modulus of both terms
in (\ref{110}) are approximately equal. Hence the terms in (\ref{110})
could even  cancel each other.\footnote{ For detailed calculations 
 of the probability of the 
$0\nu\beta\beta$-decay of different nuclei in the case of the neutrino mass hierarchy see \cite{BilFS}} From
(\ref{46}),  (\ref{47}) and (\ref{110}) for the upper bound of
the effective Majorana mass we find the value 
\be 
|m_{\beta\beta}| \leq 6.6 \cdot 10^{-3}~\rm{eV}.\label{111}
\ee
Thus, in the case of the neutrino mass hierarchy {\em upper bound of  $|m_{\beta\beta}|$} 
is smaller that the best expected sensitivity of the future  experiments on the search for 
$0\nu\beta\beta$-decay.

II. Inverted hierarchy of the neutrino masses 

 \be
m_{3} \ll m_{1} <m_{2}. \label{112} 
\ee 
For neutrino masses $m_{1,2}$
in the case of the inverted hierarchy we have
\be
m_{1}\simeq \sqrt{ |\Delta m^{2}_{13}|};~
m_{2}\simeq\sqrt{ |\Delta m^{2}_{13}|} (1 +\frac{\Delta m^{2}_{12}}{2\, |\Delta m^{2}_{13}|})\simeq\sqrt{ |\Delta m^{2}_{\rm{13}}|}.
\label{113}
\ee
The lightest neutrino mass is small:
$m_{3}\ll \sqrt{ |\Delta m^{2}_{\rm{13}}|}$.

Neglecting the contribution of the 
small term $m_{3}\, \sin^{2}\theta_{13}$, for the
effective Majorana mass we obtain the following expression
\be
|m_{\beta\beta}|\simeq \sqrt{| \Delta m^{2}_{\rm{13}}|}\,~ (1-\sin^{2}
2\,\theta_{\rm{12}}\,\sin^{2}\alpha_{12})^{\frac{1}{2}},\label{114}
\ee
where the only unknown parameter  is $\sin^{2}\alpha_{12}$. 
For the effective Majorana mass we obtain the following range
of values of the effective Majorana mass
\be 
\cos  2\,\theta_{\rm{12}} \,\sqrt{ |\Delta m^{2}_{\rm{13}}|}
\leq |m_{\beta\beta}| \leq\sqrt{ |\Delta m^{2}_{\rm{13}}|}
\label{115}
\ee
From analysis of the solar-KamLAND neutrino oscillation data it was found that
$\theta_{\rm{12}}<\pi/4$ (see (\ref{47})). Thus, the lower bound of
the effective Majorana mass in the case of the inverted mass hierarchy is different from zero.
From (\ref{46}), (\ref{47})  and   (\ref{115})   we find the following 90 \% CL range
\be 
0.9\cdot 10^{-2}\leq |m_{\beta\beta}|\leq 5.8\cdot
10^{-2}~\rm{eV}\label{116}
\ee
The anticipated sensitivities to $|m_{\beta\beta}|$ of the future
experiments on the search for $0\nu\beta\beta$ are in the range
(\ref{116}). Thus, next generation of the $0\nu\beta\beta$-
experiments will probe the inverted hierarchy of the neutrino
masses.

III. Quasi-degenerate neutrino mass spectrum

If the lightest neutrino mass satisfies inequality
\be 
m_{1} \gg \sqrt{ \Delta m^{2}_{\rm{23}}}~~~ (m_{{3}}\gg
\sqrt{ |\Delta m^{2}_{\rm{13}}|})
\label{117}
\ee
neutrino mass spectrum is practically degenerate
\be 
m_{1}\simeq m_{2}\simeq m_{3} 
\label{118}
\ee
For the effective Majorana mass 
we have in this case
\be 
|m_{\beta\beta}|\simeq m_{0}\, (1-\sin^{2}
2\,\theta_{\rm{12}}\,\sin^{2}\alpha_{12})^{\frac{1}{2}},
\label{119}
\ee
where $m_{0}$ is  the common neutrino mass. From this expression we have
\be 
\cos  2\,\theta_{\rm{12}} \,m_{0}
\leq |m_{\beta\beta}| \leq m_{0}
\label{120}
\ee
The common mass $m_{0}$ can be measured in tritium $\beta$-decay experiments. 
The expected sensitivity of 
the future KATRIN experiment \cite{Katrin}  is 
\be
m_{0}\simeq 0.2~\rm{ eV}.
\label{121}
\ee
In the case of the  quasi-degenerate neutrino mass spectrum
an information about the value of $m_{0}$ can be also obtained from the $0\nu\beta\beta$-decay 
data. From  (\ref{47}) and (\ref{119}) we find 
\be
|m_{\beta\beta}|\leq m_{0}\leq  4.4\, |m_{\beta\beta}|
\label{122}
\ee
The three neutrino mass spectra, we have considered,  correspond to different mechanisms of neutrino mass generation (see \cite{Feruglio}).
Masses of quarks and charged leptons satisfy hierarchy of the type  (\ref{108}). Hierarchy of neutrino masses is a typical feature of GUT models (like SO(10)) in which quarks and leptons are unified. Inverted spectrum and quasi-degenerate spectrum require specific symmetries of the neutrino mass matrix.

We will discuss now briefly possibilities to reveal CP violation in the case of the Majorana neutrinos.
(for the detailed discussion see \cite{PasPetSch})
We will consider only inverted and quasi-degenerate neutrino mass spectra. If CP is conserved 
in the lepton sector, neutrino mixing matrix satisfies the condition (see section 3)
\be
U_{li}= \eta_{i}\, U^{*}_{li},
\label{123}
\ee
where $\eta_{i}=\pm i$ is the the CP parity of $\nu_{i}$. Let us present $\eta_{i}$ in the form 
\be 
\eta_{i}=e^{i\frac{\pi}{2}\,\rho_{i}} ,
\label{124} 
\ee 
where
$\rho_{i}=\pm 1$. From (\ref{123}) and (\ref{124}) 
we find  
\be
e^{2i\,\alpha_{i}}=e^{\frac{i\pi}{2}\,\rho_{i}} 
\label{125} 
\ee 
In the quark sector violation of the CP invariance can be revealed
\begin{itemize}
\item
through the observation of the processes which are forbidden if CP is conserved
\item
through
the measurement of CP-odd asymmetries which are equal to zero in the case of  CP conservation.
\end{itemize}
The observation in 1964
of the decay $K_{L}\to \pi^{+}\pi^{-}$ \cite{Cronin}, which is forbidden if  CP is conserved, marked the discovery
of the CP violation in the quark sector.
The example of the second type of measurement was 
high precision measurement at the B-factories of
the CP-asymmetry in the decays $B^{0}(\bar B^{0}) \to J/ \Psi K_{S} $ \cite{Babar,Belle}.

The violation of CP invariance in the case of Majorana neutrinos can be revealed only through the
{\em measurement of
the value of the effective Majorana mass}.
From (\ref{114}) and (\ref{119}) we find 
\be 
\sin^{2}\alpha_{12}=\frac{1}{\sin^{2}
2\,\theta_{12}}~(1-\frac{|m_{\beta\beta}|^{2}}{| \Delta
m^{2}_{13}|})
\label{126}
\ee
and 
\be 
\sin^{2}\alpha_{12}=\frac{1}{\sin^{2}
2\,\theta_{12}}~(1-\frac{|m_{\beta\beta}|^{2}}{m^{2}_{0}})
\label{127}
\ee
correspondingly, for inverted hierarchy and quasi-degenerate spectrum.

In the case of the CP conservation from
(\ref{125}) we have 
\be
\sin^{2}\alpha_{12}= \sin^{2}\frac{\pi}{4} (\rho_{2}-\rho_{1})
\label{128} 
\ee 
Thus if CP is conserved parameter  $\sin^{2}\alpha_{21}$ takes {\em two values}
\be 
\sin^{2}\alpha_{21}=0 ~(\rho_{2}=\rho_{1});~
\sin^{2}\alpha_{21}=1~(\rho_{2}=-\rho_{1}). 
\label{129} 
\ee 
These values correspond to the upper and lower bounds in (\ref{115}) and 
(\ref{120}).

In the case of inverted hierarchy the determination of the CP parameter $\sin^{2}\alpha_{21}$
requires only measurement of the effective Majorana mass.\footnote{ Notice that parameters  $|\Delta m^{2}_{13}|$ and 
$\sin^{2}
2\,\theta_{12}$ will be known from the data of future experiments with much better accuracy than today.
In T2K experiment the
parameter $|\Delta m^{2}_{31}|$ will be measured with an accuracy $
5\%$ \cite{T2K}. In future solar neutrino experiments, in which $pp$
neutrinos will be detected, an accuracy $\simeq 5\%$ in the
measurement of the parameter $\sin^{2}\,\theta_{12} $  is planned to
be reached \cite{futsolar}.}
In the case of the quasi-degenerate spectrum 
common mass $m_{0}$ must be also known.

The  major uncertainty in the determination of the effective
Majorana mass $|m_{\beta\beta}|$ is connected with 
nuclear matrix elements. We will discuss this problem in the next section. Even if the problem of NME 
will be solved, CP violation in the case of the Majorana mixing can be revealed only if half-lives of $0\nu\beta\beta$-decay
will be measured with high accuracy \cite{PasPetSch}. For illustration let us consider 
in the case of the inverted hierarchy of neutrino masses
maximal CP violation 
($\alpha_{21}=\frac{\pi}{4}$)  and CP conservation. For effective Majorana mass we have, correspondingly.
\bea 
|m_{\beta\beta}|_{\rm{max}}&=& 0.76\, \sqrt{|\Delta m^{2}_{31}|}\nonumber\\
|m_{\beta\beta}|_{\rm{CP}1}&=&\sqrt{|\Delta m^{2}_{31}|};\,~~
|m_{\beta\beta}|_{\rm{CP}2}=0.38\, \sqrt{|\Delta m^{2}_{31}|}
\label{130} 
\eea 
Apparently,  it will be easier to exclude one of the CP values of the effective Majorana mass
than to decide whether CP is violated or not.

\subsection{On nuclear matrix elements}

Effective Majorana mass is not directly measurable quantity.
From experimental data only  {\em the product} of the
effective Majorana mass and nuclear matrix element  can be
obtained.  In order to determine $m_{\beta\beta}$ we must know nuclear matrix elements.

 The calculation of NME is a complicated nuclear problem (see reviews\cite{FSimCivSuh}). 
 NME is  the matrix element of  an integral which includes the T-product of two hadronic charged weak currents and neutrino
propagator. Many intermediate nuclear states must be taken into
account in  calculations.

Two different approaches 
are used for the calculation of NME:
Nuclear Shell Model (NSM) and Quasiparticle Random Phase
Approximation (QRPA). In literature exist many  QRPA-based  models.
Different calculations of NME for the same nuclear transition differ by factor 2-3
and more. In such a situation it is important to find a possibility 
to test NME calculations.

 We will discuss here  such a possibility \cite{BilGri}  which is based on the 
factorization
property of the matrix element of $0\nu\beta\beta$-decay (see section 7).

As we  have discussed  in the beginning of this section 
in several future experiments on the search for 
$0\nu\beta\beta$-decay of {\em different nuclei}  comparable sensitivities to $|m_{\beta\beta}|$ are expected. 
Thus,
if $0\nu\beta\beta$-decay of one nuclei  will be discovered in a future experiment 
it is quite probable that the 
decay will be observed also in other  experiments with different  nuclei.

A model of the calculation of NME is compatible with data only in the case if the value of the effective 
Majorana mass, determined from the results of experiments on the detection of 
$0\nu\beta\beta$-decay of different nuclei , is the same. From this requirement 
for a model $M$ we obtain the following relation 
\be
R_{M}(i;k)=\frac{T^{0\,\nu}_{1/2}(A_{k},Z_{k})}{T^{0\,\nu}_{1/2}(A_{i},Z_{i})},
\label{131} 
\ee
where 
\be
R_{M}(i;k)=\left(\frac{|M^{0\,\nu}(A_{i},Z_{i})|^{2}}
{|M^{0\,\nu}(A_{k},Z_{k})|^{2}}\right)_{M}~
\frac{G^{0\,\nu}(E^{i}_{0},Z_{i})}{G^{0\,\nu}(E^{k}_{0},Z_{k})}.
\label{132}
\ee
For illustration we will consider three 
latest models of NME calculations:
\begin{itemize}
\item ($M_{1}$) Shell Model \cite{NSM} 

\item ($M_{2}$) QRPA model \cite{RodFed}(QRPA parameter $g_{pp}$ 
is  determined from the data of the experiments on the measurement of half-
lives of the $2\nu\beta\beta$-decay.)

\item ($M_{3}$) QRPA model \cite{Civitarese}(QRPA parameters 
are determined from the $\beta$-decay data of nearby nuclei)
\end{itemize}
The results of the calculation of the ratios 
$R(^{130} \rm{Te};^{76} \rm{Ge})$, $R(^{136} \rm{Xe};^{76} \rm{Ge}) $
and $R(^{130} \rm{Te};^{136} \rm{Xe})$,
are presented in the Table I.
\newpage
\begin{center}
 Table I
\end{center}
\begin{center}
The  ratios $R_{M}(i;k)$, calculated in three recent NME models 
$M_{1}$ \cite{NSM}, $M_{2}$ \cite{RodFed}, $M_{3}$\cite{Civitarese}.
\end{center}
\begin{center}
\begin{tabular}{|c|ccc|}
\hline
&
$M_{1}$ & $M_{2}$ & $M_{3}$
\\
\hline & & & \\$R(^{130} \rm{Te};^{76} \rm{Ge})$ 
&4.08 & 2.65 & 7.76
\\ & & &  \\
\hline$R(^{76} \rm{Ge};^{136} \rm{Xe})$ & 0.56 & 0.80 & 0.07
\\ & & & \\
\hline$R(^{130} \rm{Te};^{136} \rm{Xe})$ & 2.29 & 2.11 & 0.53
\\ & & & \\
\end{tabular}
\end{center}
Because we use the factorization property of the matrix elements of 
the $0\nu\beta\beta$-decay we can compare with experimental data only ratios of NME. 
However, for the determination of the effective Majorana mass we need to know the value of NME. It could happen that for specific nuclei the ratios of NME calculated in different models are 
practically the same, in spite the values of NME being different. We see from the Table I that 
$R(^{130} \rm{Te};^{136} \rm{Xe})$ for the models $M_{1}$ and $M_{2}$ differ less than 10 \%.
However, the values of
the effective Majorana mass which can be obtained with the help of
these two models are quite different:
\be
|m_{\beta\beta}|^{2}_{M_{1}}= 1.90\,~ |m_{\beta\beta}|^{2}_{M_{2}}
\label{133}
\ee
It is evident from the Table I that   the observation of the $0\nu\beta\beta$-decay of 
$^{130} \rm{Te}$ and $^{76} \rm{Ge}$ could easily allow to decide which of the three 
considered models  is compatible with data (if any).
Generally, we can conclude that the observation of
$0\nu\beta\beta$-decay of three (or more) nuclei would be an
important tool  for the test of the models of NME calculation and
for the determination of the value of the effective Majorana mass.

\section{Conclusion}

Discovery of neutrino oscillations driven by small neutrino masses and neutrino mixing 
took many years of enormous 
efforts of many physicists.
From the point of view of the modern physics it is quite natural that neutrinos have masses.\footnote{However, at the sixties and at the seventies when
first ideas of neutrino oscillations  and mixing were proposed \cite{BP,MNS}
there was a common opinion, based on the success 
of the two-component neutrino theory, that neutrinos are massless particles.} 
The puzzling feature of 
the discovered phenomenon is extreme smallness of neutrino masses.
The most natural see-saw explanation of the smallness of neutrino masses 
requires  a violation of the total lepton number  
and Majorana neutrinos.

Neutrino oscillations are very sensitive to small neutrino mass-squared differences.
However, neutrino oscillations and transition of neutrinos in matter are 
blind to the nature of $\nu_{i}$. 
The Majorana nature of $\nu_{i}$ can be established only through the observation  of processes in which 
total lepton number is not conserved. It is a 
general feature of the standard weak interaction that 
the probabilities of such processes are very strongly suppressed.
Investigation of the 
$0\nu\beta\beta$-decay  of even-even nuclei is the most sensitive probe of  the Majorana nature of
neutrinos . Experiments on the search for $0\nu\beta\beta$-decay have very good chances to 
solve the fundamental problem of the nature of 
neutrinos with definite masses. 

The observation of $0\nu\beta\beta$-decay would be a 
direct proof that $\nu_{i}$ are Majorana particles.
From the study of this decay very important quantity,
 the effective Majorana mass $|m_{\beta\beta}|$,  can be inferred.
Determination of $|m_{\beta\beta}|$ would allow to obtain an information about 
the pattern of the neutrino mass
spectrum and lightest neutrino mass. However, to determine $|m_{\beta\beta}|$ from experimental data we need to know nuclear matrix elements.
The calculations of NME is a challenging problem for nuclear physics. 
At the moment there is no agreement between different calculations. Further progress is definitely needed. A possible test of NME calculations was discussed here.

If $\nu_{i}$ are Majorana particles neutrino mixing matrix contain additional 
(with respect to the Dirac case) CP phases (two for the three-neutrino mixing). 
Effect of these phases can be revealed through the study of   
$0\nu\beta\beta$-decay. 
However, this can be done only after the problem of NME will be solved. The high-precision 
measurement of the half-lives of $0\nu\beta\beta$- decay  will also be  required.

We can consider four possible scenarios assuming 
that the problem of NME will be solved and expected sensitivities of future experiments 
on the search for $0\nu\beta\beta$-decay and on the search for distortion of the end-point part of 
the $\beta$-spectrum of tritium
will be reached.
\begin{enumerate}
\item In the KATRIN experiment effect of the neutrino mass is detected, but  $0\nu\beta\beta$-decay 
is not observed. This would mean that neutrino spectrum is quasi-degenerate and 
$\nu_{i}$ are Dirac particles.

\item Neutrinoless double $\beta$-decay with effective Majorana mass in the range
 (\ref{120}) is observed but in the  KATRIN experiment effect of the neutrino mass is not detected.
This would mean that $\nu_{i}$ are Majorana particles, spectrum is quasi-degenerate, 
but the sensitivity of the 
KATRIN experiment is not enough to see effect of neutrino mass.

\item Neutrinoless double $\beta$-decay of different nuclei
 is observed with  $|m_{\beta\beta}|$ is in the range (\ref{115}).
It would  be a proof that 
$\nu_{i}$ are Majorana particles and inverted hierarchy of neutrino masses 
is realized.
\item Neutrinoless double $\beta$-decay is not observed in the future experiments and in  
the KATRIN experiment no distortion of $\beta$-spectrum of tritium is detected. This would mean that 
either $\nu_{i}$ are Dirac particles with masses 
smaller than the sensitivity of the KATRIN experiment
or $\nu_{i}$ are Majorana particles but spectrum is hierarchical.

\end{enumerate}

In some years we will apparently know which scenario was prepared for us by
the Nature.

It is my pleasure to acknowledge the support of  the Italien Program  ``Rientro dei cervelli''.

\end{document}